# Discrete-Time Poles and Dynamics of Discontinuous Mode Boost and Buck Converters Under Various Control Schemes

Chung-Chieh Fang*

**Abstract**

Nonlinear systems, such as switching DC-DC boost or buck converters, have rich dynamics. A simple one-dimensional discrete-time model is used to analyze the boost or buck converter in discontinuous conduction mode. Seven different control schemes (open-loop power stage, voltage mode control, current mode control, constant power load, constant current load, constant-on-time control, and boundary conduction mode) are analyzed systematically. The linearized dynamics is obtained simply by taking partial derivatives with respect to dynamic variables. In the discrete-time model, there is only a single pole and no zero. The single closed-loop pole is a linear combination of three terms: the open-loop pole, a term due to the control scheme, and a term due to the non-resistive load. Even with a single pole, the phase response of the discrete-time model can go beyond -90 degrees as in the two-pole average models. In the boost converter with a resistive load under current mode control, adding the compensating ramp has no effect on the pole location. Increasing the ramp slope decreases the DC gain of control-to-output transfer function and increases the audio-susceptibility. Similar analysis is applied to the buck converter with a non-resistive load or variable switching frequency. The derived dynamics agrees closely with the exact switching model and the past research results.

**KEY WORDS:** Nonlinear system, DC-DC power conversion, discrete-time model, non-resistive load, discontinuous conduction mode, small-signal analysis

*C.-C Fang is with Advanced Analog Technology, 2F, No. 17, Industry E. 2nd Rd., Hsinchu 300, Taiwan, Tel: +886-3-5633125 ext 3612, Email: fangcc3@yahoo.com

# Contents

Table 1: Summary of seven schemes analyzed.

|  | **Scheme** |
|---|---|
| **Part I.** | **Resistive Load (Fixed Switching Frequency)** (summarized in Table 2) |
| $\mathbb{S}_1$ | Open-loop power stage |
| $\mathbb{S}_2$ | Voltage mode control (VMC) |
| $\mathbb{S}_3$ | Current mode control (CMC) |
| **Part II.** | **Non-Resistive Load (Fixed Switching Frequency)** (summarized in Table 3) |
| $\mathbb{S}_4$ | Power stage/VMC/CMC with constant power load (CPL) |
| $\mathbb{S}_5$ | Power stage/VMC/CMC with constant current load (CCL) |
| **Part III.** | **Part III. Variable Switching Frequency Control** (summarized in Table 4) |
| $\mathbb{S}_6$ | Valley voltage constant-on-time control (V-COTC) |
| $\mathbb{S}_7$ | Boundary Conduction Mode (BCM) |

# 1 Introduction

Nonlinear systems, such as switching DC-DC boost or buck converters, have rich dynamics. Many efforts have been made in the past three decades to analyze the boost converter *power stage* in discontinuous conduction mode (DCM) based on average models [1–8]. Fewer efforts [9–11] have been made to model the boost converter under *current mode control* (CMC) in DCM. The analysis of the DCM is generally believed to be complex because DCM has three stages in a switching cycle. Combination of CMC and DCM further increases the complexity. Adding a non-resistive load also increases the complexity [12]. This paper presents an alternative and accurate modeling in addition to the average models.

Similar modeling approach has been applied to the fixed-switching-frequency buck converter with a resistive load [13]. Compared with [13], this paper makes three additional extensions. First, it extends to the boost converter. Second, it extends to the *non-resistive* load case. Third, it extends to *variable* switching frequency control. This paper focuses on the boost converter.

In [14], a discrete-time model for the boost converter *power stage* in DCM is proposed. The model accurately predicts subharmonic oscillation in a boost converter with proportional voltage feedback [15]. However, its potential advantage has not been fully appreciated. In this paper, the discrete-time model is applied to analyze the boost and buck converters in seven different schemes in a unified way (summarized in Table 1): open-loop power stage, voltage mode control (VMC), current mode control, constant power load (CPL), constant current load (CCL), constant-on-time control (COTC), and boundary conduction mode (BCM).

In the past, the analysis of these different schemes were reported in separate references [1, 3–7, 9–12, 16–18], instead of in a single reference as this paper. Here, the linearized dynamics of the *discrete-time* model is obtained simply by taking partial derivatives with respect to dynamic variables. In the discrete-time model, the pole will be shown to have a simpler expression and is a linear combination of three terms: the open-loop pole, a term due to the control scheme, and a term due to the non-resistive load. The discrete-time model provides a simpler *alternative* to design or analyze the converter different from the circuit-averaging approach [10]. The same methodology developed here can be extended to analyze other types of converters [13, 19], or applications, such as power factor correction and *digital* control of DC-DC converters *directly* based on discrete-time dynamics. For example, as shown in this paper, similar analysis can be readily extended to analyze the converter with a *non-resistive* load.

This paper presents theoretical analysis of already experimentally observed phenomena in [5, 9, 12, 15–17, 20, 21]. Seven simulation examples based on the *exact* switching model have been made.

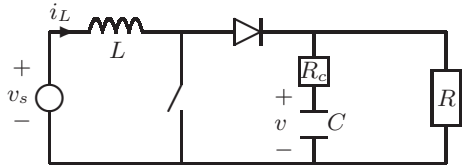

Figure 1: A boost converter power stage.

All the obtained results agree coherently with the past observations, and they are verified by the exact switching model. The analysis of the seven different schemes is presented next. For each scheme, the flow of analysis is as follows.

1. Identify the *dynamic* variables.
2. Identify the *feedback* variables.
3. Derive the switching *constraint* (*when* the switch is turned on/off).
4. Derive the large-signal dynamics.
5. Using partial derivatives, derive the small-signal (linearized) dynamics.
6. Determine the pole, control-to-output and audio-susceptibility frequency responses.

This paper has three parts. In Part I, the boost converter with resistive load is analyzed, and similar results [13] for the buck converter are reviewed. In Part II, pole shifting due to *non-resistive* load is analyzed. In Part III, *variable* frequency control is analyzed. At the end of each part, the key results are summarized in a table.

# Part I
# Open/Closed-Loop Dynamics, VMC, and CMC

## 2  Scheme One ($\mathbb{S}_1$): Open-Loop Power Stage

Consider a boost converter power stage (Fig. 1) with a switching frequency $f_s$ and a switching period $T = 1/f_s$. Let $\omega_s = 2\pi f_s$. Denote the source voltage as $v_s$, the capacitor voltage as $v$, the inductance as $L$, the capacitance as $C$, and the equivalent series resistance (ESR) as $R_c$. The load, either resistive or non-resistive, has a steady-state effective resistance $R$. If the load is non-resistive, the dynamic resistance in the $n$-th cycle is denoted as $R_n$.

### 2.1  Nonlinear Discrete-Time Model

In DCM, there are three stages in the switching period. Let the durations of the first and the second stages be $DT$ and $D_2T$, respectively. The inductor current $i_L$ is zero in the third stage, and one discrete-time pole is zero [22]. The *discrete-time* dynamics is thus one-dimensional.

Throughout the paper, to simplify the dynamics, all *continuous-time* variables are sampled at the beginning of each cycle. A subscript $n$ is used for a dynamic variable of the $n$-th cycle. For example, $D_n$ denotes the duty cycle in the $n$-th cycle. Also, trailing-edge modulation (where the switch is turned on at the beginning of each cycle) is assumed. An illustrative signal plot of $i_L$ is shown in Fig. 2. The nonlinear large-signal discrete-time model (mapping) reported in [14] is rearranged here as

$$v_{n+1} = f(v_n, v_{sn}, D_n, R_n) = (1 - \beta_n K_n)v_n + \frac{\beta_n v_{sn}^2 D_n^2}{v_n - v_{sn}} \qquad (1)$$

where $K_n = 2L/R_n T$, $\beta_n = \rho_n T^2/2LC$, and $\rho_n = R_n/(R_n + R_c)$. The model dynamics is shown in Fig. 3. Note that $K_n$, $\beta_n$, and $\rho_n$ are dimensionless variables. Also note that $R_n$, $K_n$, $\beta_n$, and $\rho_n$ are dynamic variables (varying in each cycle) if the load is non-resistive. If the load is resistive, they are constant and denoted as $R$, $K$, $\beta$, and $\rho$, respectively. The short notation $v_{sn}$, instead of $v_{s,n}$, is used for brevity. This applies to other variables. For $R_c = 0$, $\rho_n = 1$. Here, under fixed switching frequency, $T$ is constant. In Sec. III, under variable switching frequency, and the switching period is a dynamic variable, denoted as $T_n$.

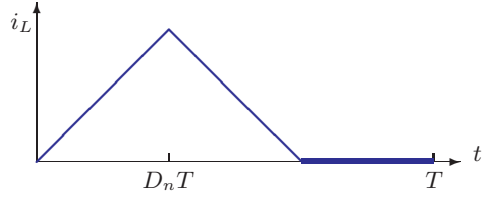

Figure 2: An illustrative signal plot of $i_L$ in each cycle for DCM.

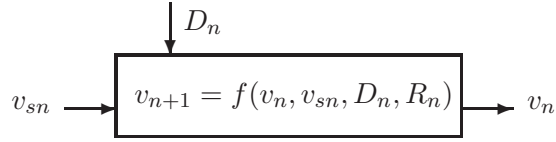

Figure 3: Open-loop power stage large-signal dynamics.

## 2.2 Steady-State (Fixed-Point) Analysis

In steady state, let (fixed-points) $D_n = D$, $R_n = R$, $K_n = K = 2L/RT$, $\rho_n = \rho = R/(R + R_c)$, $\beta_n = \beta = \rho T^2/2LC$, $v_{sn} = v_s$ and $v_{n+1} = v_n = v = Mv_s$, where $M$ is the conversion ratio [10]. Then (1) leads to a steady-state equation,

$$M^2 - M - \frac{D^2}{K} = 0 \qquad (2)$$

which is a quadratic equation of $M$ and has *two* solutions. Ignoring the negative solution, one has the same result as in [10, p. 124],

$$M = \frac{1 + \sqrt{1 + 4D^2/K}}{2} \qquad (3)$$

Assume ESR is small. By simple algebra based on the steady-state inductor current slopes, one has

$$\frac{D_2}{D} = \frac{v_s}{v_s - v} = \frac{1}{1 - M} \tag{4}$$

Using (3) and (4), one has $D_2 = KM/D$ and $D = (M-1)D_2 = \sqrt{KM(M-1)}$. These equations greatly simplify the linearized dynamics and are used throughout the paper.

## 2.3 Limitation on the Input Space

The nonlinear dynamics (1) is derived under the assumption that the converter operates in DCM, not all inputs $(v_n, v_{sn}, D_n, R_n) = (v, v_s, D, R)$ are legitimate, unless an additional CCM model is included. An input $(v, v_s, D, R)$ which makes the converter leave DCM is illegitimate. Assume that the variation of the capacitor voltage $v$ is small within a switching period $T$, then $D_2/D = v_s/(v - v_s) = 1/(M-1)$. The converter operating within DCM requires $D + D_2 = Dv/(v - v_s) < 1$, which leads to the following limitation on the input space $(v, v_s, D, R)$ for the mapping (1) to be legitimate:

$$\frac{1}{1-D} < \frac{v}{v_s} \tag{5}$$

## 2.4 Linearized Open-Loop Dynamics

A hat ˆ is used to denote small perturbations (e.g., $\hat{v}_n = v_n - v$ and $\hat{D}_n = D_n - D$). For a resistive load, $\hat{R}_n = R - R = 0$. The effect of a non-resistive load will be discussed later. The linearized open-loop dynamics is

$$\hat{v}_{n+1} = \frac{\partial f}{\partial v_n}\hat{v}_n + \frac{\partial f}{\partial v_{sn}}\hat{v}_{sn} + \frac{\partial f}{\partial D_n}\hat{D}_n \tag{6}$$

$$= [1 - \frac{\rho T}{RC}(\frac{2M-1}{M-1})]\hat{v}_n + \frac{\rho TM}{RC}(\frac{2M-1}{M-1})\hat{v}_{sn} + [\frac{2\rho TM v_s}{RCD}]\hat{D}_n \tag{7}$$

$$:= p_0 \hat{v}_n + \Gamma_{s0}\hat{v}_{sn} + \Gamma_{c0}\hat{D}_n \tag{8}$$

where the open-loop pole is

$$p_0 = 1 - \frac{\rho T}{RC}(\frac{2M-1}{M-1}) \tag{9}$$

The converter is stable if $|p_0| < 1$. Saddle-node bifurcation [23] occurs when $p_0 = 1$, and subharmonic oscillation (period-doubling bifurcation) occurs when $p_0 = -1$.

In [10, p. 427], the *negative continuous-time* pole is $\omega_p = (2M-1)/RC(1-M)$, equivalent (through a mapping) to the discrete-time pole $p_0 \approx e^{\omega_p T}$ if $T \ll RC$ and $R_c \ll R$. Thus, given a discrete-time pole (9), one can easily obtains its corresponding continuous-time pole.

## 2.5 Agreement with the Exact Switching Model

In [24], based on the exact switching model, the *exact* value of the discrete-time pole $p_0$ is obtained,

$$p_0 = e^{-\rho\omega_c(T-\frac{d_2-d_1}{2})}e^{-\rho\omega_l(\frac{d_2-d_1}{2})}((\frac{\omega_l - \omega_c}{2\omega})\sin(\rho\omega(d_2-d_1)) + \cos(\rho\omega(d_2-d_1))) \tag{10}$$

where $d_1 = DT$, $d_2 = (D_2 + D)T$, and

$$\begin{aligned}\omega_c &= \frac{1}{RC} & \omega_l &= \frac{R_c}{L} \\ \omega &= \sqrt{\omega_0^2 - (\frac{\omega_c - \omega_l}{2})^2} \approx \omega_0 & \omega_0 &= \frac{1}{\sqrt{LC}}\end{aligned} \tag{11}$$

For a small $\theta$, $e^\theta \approx 1 + \theta$, $\sin(\theta) \approx \theta$ and $\cos(\theta) \approx 1 - \theta^2/2$, then the exact discrete-time pole (10) becomes (7). This shows that the derived linearized model (8), although based on the approximate nonlinear model (1), is close to the *exact* switching model.

## 2.6 Open-Loop Frequency Responses

The output voltage is close to $\rho v$. In the power stage, $D_n$ is the control variable to control the output voltage. Given the dynamics (8), the open-loop control-to-output transfer function is

$$T_{oc0}(z) := \frac{\rho \hat{v}(z)}{\hat{D}(z)} = \frac{\rho \Gamma_{c0}}{z - p_0} \qquad (12)$$

As discussed above, $D = \sqrt{KM(M-1)}$. The DC gain, agreed with [10, p. 427], is

$$T_{oc0}(1) = \frac{\rho \Gamma_{c0}}{1 - p_0} = \frac{2\rho v}{D}\left(\frac{M-1}{2M-1}\right) = \frac{2\rho v_s}{K}\left(\frac{D}{2M-1}\right) \qquad (13)$$

Similarly, the open-loop audio-susceptibility (source-to-output transfer function) is

$$T_{os0}(z) := \frac{\rho \hat{v}(z)}{\hat{v}_s(z)} = \frac{\rho \Gamma_{s0}}{z - p_0} \qquad (14)$$

and the DC gain of audio-susceptibility is

$$T_{os0}(1) = \frac{\rho \Gamma_{s0}}{1 - p_0} = \rho M \qquad (15)$$

agreed with [2].

Given a transfer function in the $z$-domain, say $T(z)$, its DC gain is $T(1)$, and its effective frequency response is $T(e^{j\omega T})$, which is valid in the frequency range $|\omega| < \omega_s/2$ (half the switching frequency). Different from a single-pole continuous-time system, in which the phase response cannot go beyond -90 degrees, a single-pole discrete-time system has phase response beyond -90 degrees [22], giving similar results as in *two*-pole average models [3–5] as shown in the next example.

**Example 1.** (*The frequency response of the discrete-time model agrees with the experimental results reported in [5].*) Consider a boost converter power stage from [5] with parameters $f_s = 100$ kHz, $v_s = 5$ V, $R = 20$ Ω, $L = 5$ μH, $C = 40$ μF, $R_c = 0$, and $D = 0.7$.

The pole from (7) is 0.9703. The exact pole from (10) based on the exact switching model [24] is 0.9707. Both agree closely. Throughout the paper, the *exact switching model* means the circuit as in Fig. 1 with the ideal switch, where the *exact* switching instants depend on the particular control scheme. Simulation based on the exact switching model is expected to be accurate as other circuit simulators such SIMPLIS, PSIM and SABER.

The control-to-output frequency response of the discrete-time model (12) is shown in Fig. 4, compared with that of the average model in [5]. The frequency response of the discrete-time model matches well with the experimental data (reproduced and marked as * in Fig. 4) reported in [5] based on SABER simulation. This example shows that the discrete-time model, even though with only one pole and no zero, still gives accurate frequency responses.

Here, both the discrete-time model and the average model in [5] agrees closely with the exact switching model. In this example, the average model has good agreement because the discrete-time pole here is *real positive*. With additional feedback as discussed later, the closed-loop discrete-time pole may be *real negative*, and the converter is oscillatory [13]. In that case, the discrete-time model would give more accurate results than the average model (as shown in Example 2 with voltage feedback). In the average model of the boost converter, if the ESR zero is located between the two poles in the complex plane, the root loci of poles would remain on the real axis, and the converter with any feedback gain is not oscillatory. However, if the ESR zero is located to the left of the high-frequency pole in the complex plane, based on the root locus, the converter may have complex poles, but the oscillation frequency is not subharmonic (contradicting to simulations or real circuit experiments shown in Example 2 or [15]). Therefore, the average models are accurate only in some conditions (when the discrete-time pole is *real positive*) as reported in [13], whereas the discrete model does not have such a limitation. □

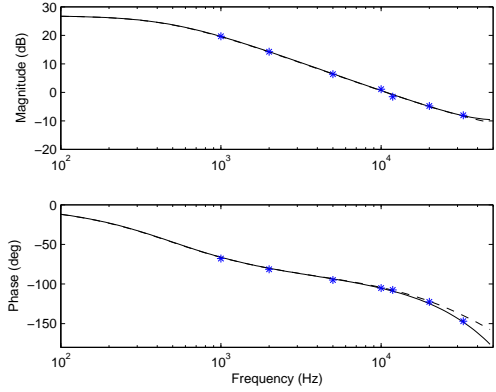

Figure 4: Control-to-output frequency responses of discrete-time model (solid line), average model [5] (dashed line) and experimental data marked as *.

## 3 General Closed-Loop Dynamics with Non-Resistive Load

### 3.1 The Pole is a Linear Combination of Three Terms

For a closed-loop converter, another switching constraint associated with the duty cycle is placed on the power stage dynamics (1). Generally, the constraint can be represented directly in terms of the duty cycle as a function of other variables,

$$D_n = D(v_n, v_{sn}, v_{cn}, R_n) \tag{16}$$

where, in VMC, the control variable $v_{cn}$ controls the output voltage; while in CMC, the control variable $v_{cn}$ controls the peak inductor current. The closed-loop dynamics is shown in Fig. 5.

Generally, the dynamic load $R_n$ can be represented as a function of the capacitor voltage $v_n$, $R_n = R(v_n)$. For general cases about other switching constraints or other load representations, similar dynamics can be derived and are omitted to save space.

From (1) and (16), the linearized closed-loop dynamics is

$$\begin{align}
\hat{v}_{n+1} &= [\frac{\partial f}{\partial v_n} + \frac{\partial f}{\partial D_n}\frac{\partial D}{\partial v_n} + \frac{\partial f}{\partial R_n}\frac{\partial R}{\partial v_n}]\hat{v}_n + [\frac{\partial f}{\partial v_{sn}} + \frac{\partial f}{\partial D_n}\frac{\partial D}{\partial v_{sn}}]\hat{v}_{sn} + [\frac{\partial f}{\partial D_n}\frac{\partial D}{\partial v_{cn}}]\hat{v}_{cn} \tag{17} \\
&:= p\hat{v}_n + \Gamma_s\hat{v}_{sn} + \Gamma_c\hat{v}_{cn} \tag{18}
\end{align}$$

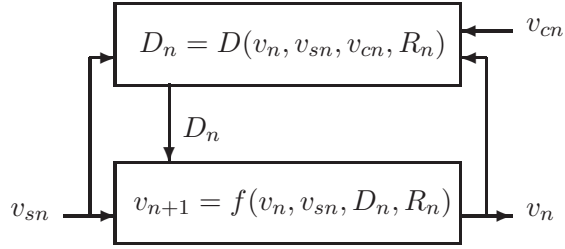

Figure 5: Closed-loop large-signal dynamics.

where $p$ is the *closed-loop* pole and can be expressed as a combination of three terms:

$$p = \frac{\partial f}{\partial v_n} + \frac{\partial f}{\partial D_n}\frac{\partial D}{\partial v_n} + \frac{\partial f}{\partial R_n}\frac{\partial R}{\partial v_n} \quad (19)$$
$$:= p_0 + \Delta p_c + \Delta p_l \quad (20)$$

where $\Delta p_c = (\partial f/\partial D_n)(\partial D/\partial v_n)$ denotes a pole shifting due to the closed-loop *control* scheme, and $\Delta p_l = (\partial f/\partial R_n)(\partial R/\partial v_n)$ denotes a pole shifting due to the *non-resistive* load. Note that the term $\Delta p_c$ depends on the control scheme, and may be converter-dependent, whereas the $\Delta p_l$ is generally converter-independent. If the load is purely resistive, one has $\partial R/\partial v_n = 0$ and $\Delta p_l = 0$. Note that a *resistive* load affects the *pole* location through $R$, as shown in (9), not through $\Delta p_l$.

### 3.2 Closed-Loop Frequency Responses

Given the dynamics (18), the closed-loop control-to-output transfer function is

$$T_{oc}(z) := \frac{\rho \hat{v}(z)}{\hat{v}_c(z)} = \frac{\rho \Gamma_c}{z - p} \quad (21)$$

Similarly, the closed-loop audio-susceptibility (source-to-output transfer function) is

$$T_{os}(z) := \frac{\rho \hat{v}(z)}{\hat{v}_s(z)} = \frac{\rho \Gamma_s}{z - p} \quad (22)$$

Compared with other modeling approaches, the discrete-time modeling is simpler. Given a converter with a particular load under a particular control scheme, the discrete-time pole is just a linear combination of different terms. The dynamics for the boost converter under fixed-switching-frequency VMC or CMC with a resistive load is presented next, followed by the non-resistive load case in Sec. 7, and the variable-switching-frequency case in Sec. III.

Given a particular control scheme, the first step of analysis is to determine the switching constraint (16). Once the constraint is obtained, the closed-loop dynamics and pole can be easily obtained from (18) and (20).

## 4 Scheme Two ($\mathbb{S}_2$): Voltage Mode Control (VMC) with Resistive Load

Consider a VMC boost converter shown in Fig. 6. Assume that the output voltage variation is small *within* a cycle and $R_c$ is small. Consider a voltage feedback with a gain $g$. Let the ramp amplitude in the PWM module be $V_h$ and the reference voltage be $v_c$. The duty cycle is determined by equating the voltage loop output $g(v_c - v_n)$ to the ramp $D_n V_h$ at the switching instant, $g(v_c - v_n) = D_n V_h$. It is rearranged in terms of the duty cycle $D_n$ as a function of $v_n$,

$$D_n = D(v_n) \approx \frac{g(v_c - v_n)}{V_h} \tag{23}$$

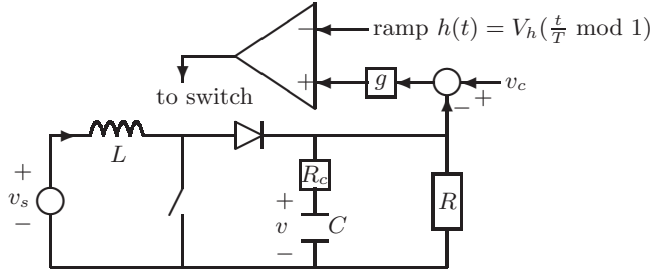

Figure 6: A boost converter under voltage-mode control.

It is a simple state feedback and the linearized closed-loop dynamics is $\hat{v}_{n+1} = p\hat{v}_n + \Gamma_{s0}\hat{v}_{sn}$, where the closed-loop pole is $p = p_0 + \Delta p_c = p_0 - g\Gamma_{c0}/V_h$. The converter is stable if $-1 < p < 1$. Since $\Delta p_c$ here is generally negative, which may shift the pole to the left and make the pole $p = p_0 + \Delta p_c$ to be negative. Subharmonic oscillation occurs when $p = p_0 - g\Gamma_{c0}/V_h < -1$, rearranged as

$$g > \frac{(p_0 + 1)V_h}{\Gamma_{c0}} \tag{24}$$

**Example 2.** (*The discrete-time model gives better prediction of gain margin than the average model.*) Consider a VMC boost converter from [15] with parameters $f_s = 3$ kHz, $v_s = 16$ V, $R = 12.5\ \Omega$, $L = 208\ \mu$H, $C = 222\ \mu$F, $R_c = 0$, and output voltage $v = 25$ V. It is shown in [15] that subharmonic oscillation occurs when $g > 0.08$ by simulation based on the exact switching model.

From (24), the critical gain (when the subharmonic oscillation occurs) is 0.076. Thus, a feedback gain greater than 0.076 is expected to be destabilizing, which agrees with the simulation result in [15] noted above. With $g = 0.076$, the closed-loop discrete-time pole is -1.08, indicating that the converter is oscillatory at the half switching frequency (9424.9 rad/s).

Next, the frequency responses of the discrete-time model and the average model are compared. It will be shown that the discrete-time model gives more accurate results. The control-to-output frequency response of the discrete-time model (21) is shown in Fig. 7, compared with that of the average model based on [5]. The gain margin of the discrete-time model is -22.4 dB (corresponding to $g = 10^{-22.4/20} = 0.076$) at the half switching frequency. The gain margin agrees with the *exact* switching model reported in [15].

In contrast, the gain margin based on the *average* model [5] is -9.28 dB (corresponding to $g = 10^{-9.28/20} = 0.3436$) at frequency 15900 rad/s, which does not accurately predict the critical gain for the subharmonic oscillation. With $g = 0.076$, the closed-loop poles of the average model are $-4508.3 \pm 7115.2i$, which are *stable* with a *transient* oscillation frequency at 7115.2 rad/s. This contradicts with the simulation that, with $g = 0.076$, the converter is unstable with a (subharmonic) oscillation frequency at 9424.9 rad/s. In this example, the discrete-time model gives more accurate results than the average model both qualitatively (about the stability) and quantitatively (about the oscillation frequency). □

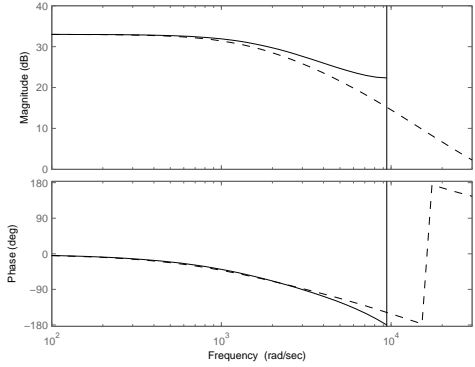

Figure 7: Control-to-output frequency responses of discrete-time model (solid line) and average model (dashed line). The gain margin -22.4 dB predicted by the discrete-time model is more accurate than the gain margin -9.28 dB predicted by the average model.

## 5 Scheme Three ($\mathbb{S}_3$): Current Mode Control (CMC) with Resistive Load

In CMC shown in Fig. 8, let the compensating ramp slope be $m_a$ and the inductor current slope in the first stage of each cycle be $m_1 = v_s/L$. The duty cycle is determined by these two slopes and the control variable $v_{cn}$ (which controls the peak inductor current):

$$D_n = D(v_{sn}, v_{cn}) = \frac{v_{cn}}{T(\frac{v_{sn}}{L} + m_a)} \tag{25}$$

This feedback control law adds a nonlinear constraint to the discrete-time dynamics (1). As noted in [9] for the *boost* converter in DCM, CMC adds feedforward from $v_s$ but adds no voltage feedback. Since $D_n$ is not a function of $v_n$, one has $\Delta p_c = (\partial f/\partial D_n)(\partial D/\partial v_n) = 0$ and $p = p_0$. The pole for CMC is the same as the open-loop power stage pole, agreed with [9]. The results for CCM and DCM are different. In CCM, CMC does add *state* feedback. In contrast, in DCM, the CMC control law (25) does not add any state feedback because the initial current at the beginning of each cycle is zero.

### 5.1 Linearized Dynamics

Let $m_c = 1 + m_a/m_1$ as in [9]. Taking partial derivative of (25) with respect to $v_{cn}$ and $v_{sn}$, the constraint (25) has linearized dynamics,

$$\hat{D}_n = \frac{\partial D}{\partial v_{cn}}\hat{v}_{cn} + \frac{\partial D}{\partial v_{sn}}\hat{v}_{sn} = [\frac{1}{Tm_1 m_c}]\hat{v}_{cn} - [\frac{D}{v_s m_c}]\hat{v}_{sn} \tag{26}$$

The closed-current-loop linearized dynamics, (7) with (26), can be simplified as

$$\hat{v}_{n+1} = p_0 \hat{v}_n + \frac{\rho T M}{RC}(\frac{2M-1}{M-1} - \frac{2}{m_c})\hat{v}_{sn} + [\frac{\rho T D}{C(M-1)m_c}]\hat{v}_{cn} \tag{27}$$

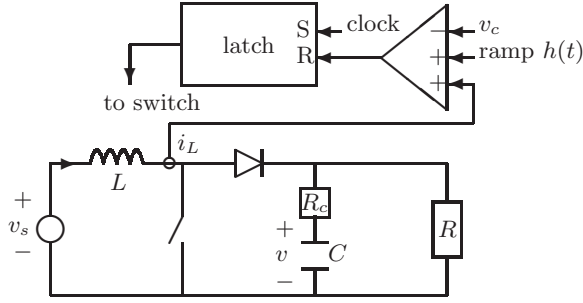

Figure 8: A boost converter under current-mode control.

## 5.2 DC Gains and the Effects of the Compensating Ramp

From (21) and (27), the DC gain of control-to-output transfer function, agreed with [9], is

$$T_{oc}(1) = \frac{\rho RD}{(2M-1)m_c} \qquad (28)$$

Compared with (13), for no compensating ramp added ($m_c = 1$), the DC gain for CMC is larger than that for the power stage if $L > Tv_s$. However, since $m_c \geq 1$, adding the ramp or increasing the ramp slope decreases the DC gain.

From (22), the DC gain of audio-susceptibility is

$$T_{os}(1) = \rho M(1 - \frac{2(M-1)}{(2M-1)m_c}) < \rho M = T_{os0}(1) \qquad (29)$$

The audio-susceptibility for CMC is smaller than that for the open-loop power stage (see (15)). However, based on (29), one has $\partial T_{os}(1)/\partial m_c > 0$, and increasing the ramp slope increases the audio-susceptibility. From (29), the DC gain of audio-susceptibility is nulled if $m_c = 1 - 1/(2M-1)$. Since $M > 1$ for the boost converter, a *negative* ramp (with $m_c < 1$) is required to null the audio-susceptibility.

Without the ramp compensation ($m_c = 1$), (29) becomes

$$T_{os}(1) = \frac{\rho M}{2M-1} \qquad (30)$$

which is close to $1/2$ for a large $M$.

Different from the CCM case, the effects of the compensating ramp for DCM are summarized. First, since the pole location in DCM is not shifted by CMC, adding the ramp also does not shift the pole and does not affect the stability. Second, increasing the ramp slope decreases the DC gain of control-to-output transfer function. Third, increasing the ramp slope increases the DC gain of audio-susceptibility. These three effects raise the question whether the ramp is needed for the boost converter in DCM. The ramp, beneficial in CCM to stabilize the current loop, may be unnecessary in DCM since the current loop itself is not oscillatory. The need of the ramp is also questioned for the buck converter in DCM [13].

**Example 3.** Consider a CMC boost converter with parameters $f_s = 700$ kHz, $v_s = 12$ V, $R = 24$ $\Omega$, $L = 1$ $\mu$H, $C = 125$ $\mu$F, $R_c = 0$, and output voltage $v = 24$ V.

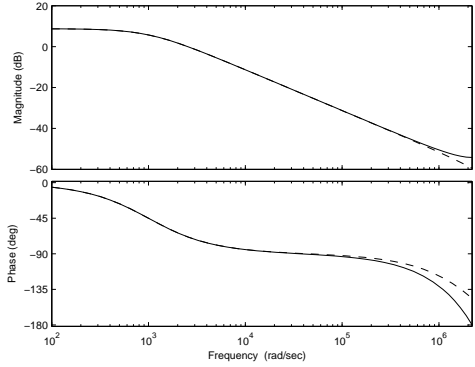

Figure 9: Control-to-output frequency responses of discrete-time model (solid line) and average model [9] (dashed line).

The control-to-output frequency response (21) is shown in Fig. 9, compared with that of the average model reported in [9] which was shown in agreement with the experimental results. Both have similar magnitude frequency responses. The discrete-time model generally has a larger phase lag close to the half switching frequency. This can be explained by the fact, in the discrete-time model, the output is measured at the start of the period ($t = nT$), while the control is exerted at $t = nT + D_n T$ (with a delay). The discrepancy can be mitigated if the output is also measured at $t = nT + D_n T$ [22]. □

## 6  Buck Converter with a Resistive Load: A Short Review

Similar results for the buck converter based on [13] are summarized here for completeness and also for comparison. From [13], the nonlinear large-signal discrete-time dynamic for the buck converter in DCM is

$$v_{n+1} = (1 - \beta K)v_n - \beta v_s D_n^2 (1 - \frac{v_s}{v_n}) \tag{31}$$

The open-loop power stage pole for the buck converter, agreed with [10, p. 427], is

$$p_0 = 1 - \frac{\rho T}{RC}(\frac{2 - M}{1 - M}) \tag{32}$$

From [13], the pole for the CMC buck converter, agreed with [9, 10], is

$$p = p_0 + \Delta p_c = p_0 + \frac{\rho T}{RCm_c}(\frac{2M}{1 - M}) = 1 - \frac{\rho T}{RC}(\frac{2 - M - \frac{2M}{m_c}}{1 - M}) \tag{33}$$

With no compensating ramp, $m_c = 1$, and the CMC pole is

$$1 - \frac{\rho T}{RC}(\frac{2 - 3M}{1 - M}) \tag{34}$$

The pole (34) is greater than 1 (unstable) for $M > 2/3$, implying occurrence of saddle-node bifurcation [13]. The possibility of instability for $M > 2/3$ was also reported in [9, 10].

Table 2: Summary for the power stage and CMC for boost and buck converters, some agreed with past research results [2, 9, 10, 24].

|  | Boost converter |  | Buck converter |  |
|---|---|---|---|---|
| **Power stage** |  |  |  |  |
| Pole, $p_0$ | $1 - \frac{\rho T}{RC}(\frac{2M-1}{M-1})$ | [24] | $1 - \frac{\rho T}{RC}(\frac{2-M}{1-M})$ | [24] |
| DC gain, $T_{oc0}(1)$ | $\frac{2\rho v}{D}(\frac{M-1}{2M-1})$ | [10] | $\frac{2\rho M v_s}{D}(\frac{1-M}{2-M})$ | [10] |
| Audio-susceptibility, $T_{os0}(1)$ | $\rho M$ | [2] | $M$ | [2] |
| **CMC** |  |  |  |  |
| Pole shifting, $\Delta p_c$ | $0$ |  | $\frac{\rho T}{RCm_c}(\frac{2M}{1-M})$ |  |
| Pole, $p = p_0 + \Delta p_c$ | $p_0$ |  | $1 - \frac{\rho T}{RC}(\frac{2-M-\frac{2M}{m_c}}{1-M})$ |  |
|  |  |  | unstable if $m_c < \frac{2M}{2-M}$ |  |
| $m_c = 1$ (no ramp) | $p_0$ |  | $1 - \frac{\rho T}{RC}(\frac{2-3M}{1-M})$ |  |
|  |  |  | unstable if $M > 2/3$ | [9] |
| DC gain, $T_{oc}(1)$ | $\frac{\rho RD}{(2M-1)m_c}$ | [9] | $\frac{R(1-M)}{Mm_c(2-M-\frac{2M}{m_c})}$ |  |
| $m_c = 1$ (no ramp) | $\frac{\rho RD}{2M-1}$ |  | $\frac{R(1-M)}{M(2-3M)}$ |  |
| Audio-susceptibility, $T_{os}(1)$ | $\rho M(1 - \frac{2(M-1)}{(2M-1)m_c})$ |  | $\frac{M(2-M-\frac{2}{m_c})}{2-M-\frac{2M}{m_c}}$ | [9] |
| $m_c = 1$ (no ramp) | $\frac{\rho M}{2M-1}$ |  | $\frac{M^2}{3M-2}$ |  |

A summary for the power stage and CMC for boost and buck converters is given in Table 2.

# Part II
# Pole Shifting due to Non-Resistive Load

## 7  Boost Converter

With a non-resistive load, either under open loop, VMC or CMC, the pole is shifted by $\Delta p_l$ as discussed in Sec. 3. Two different loads, CPL and CCL, are considered.

### 7.1  Scheme Four ($\mathbb{S}_4$): Boost Converter with a Resistive Load in Parallel with CPL

Let the load be a resistive load $R_0$ in parallel with a CPL (with a constant power $P$) as shown in Fig. 10. Assume that the output voltage variation *within* a cycle is small and ESR is also small, then the effective resistance of the CPL is close to $v_n^2/P$. The total load resistance (as a function of $v_n$) is $R_n = R(v_n) \approx R_0 \parallel (v_n^2/P) = R_0 v_n^2/(R_0 P + v_n^2)$. For $R_0 = \infty$, one has $R_n = v_n^2/P$ and the load is a pure CPL. For $P = 0$, $R_n = R_0$ and the load is a pure resistor.

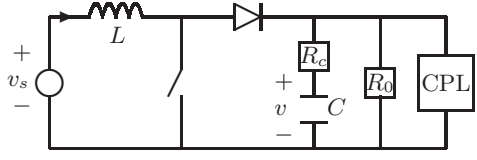

Figure 10: A boost converter power stage with CPL and resistive load $R_0$.

In steady state, the effective resistance is $R = R_0 v^2/(R_0 P + v^2)$, which leads to

$$\frac{1}{R} - \frac{1}{R_0} = \frac{P}{v^2} \tag{35}$$

From (20),

$$\Delta p_l = \frac{\partial f}{\partial R_n} \frac{\partial R}{\partial v_n} = \frac{2\rho T}{C}(\frac{1}{R} - \frac{1}{R_0}) = \frac{2\rho TP}{v^2 C} \tag{36}$$

which is independent of $R_0$. Note that, as discussed above, a resistive load such as $R_0$ affects the pole location through the effective resistance $R$ as shown in (9), not through $\Delta p_l$. Also note that, with a small $R$ or $R_0$, the converter may not operate in DCM, and the DCM analysis does not apply. For $P = 0$, one has $R = R_0$ and $\Delta p_l = 0$ because with a resistive load, the pole is not shifted by an additional term.

As noted above, for the CMC boost converter, $\Delta p_c = 0$. With CPL, the poles for the power stage and CMC are the *same*.

From (9) and (36), for either the power stage *or* CMC,

$$p = p_0 + \Delta p_l = 1 - \frac{\rho T}{RC}(\frac{2M-1}{M-1}) + \frac{2\rho TP}{v^2 C} = 1 - \frac{\rho T}{RC}(\frac{1}{M-1}) - \frac{2\rho T}{R_0 C} \tag{37}$$

When the load is a pure CPL ($R_0 = \infty$ and the total effective resistance $R = v^2/P$), for example, the converter is generally stable (with $p < 1$) and agreed with [12]. Subharmonic oscillation (with $p < -1$) may occur if $M < 1 + \rho T/2RC$ (close to 1), which is rare because another condition [10] with $M > 1/(1-D)$ for the boost converter is required in DCM.

## 7.2 Scheme Five ($\mathbb{S}_5$): Boost Converter with a Resistive Load in Parallel with CCL

### 7.2.1 General Case

Let the load be a resistive load $R_0$ in parallel with a CCL (with a constant current $I_o$) as shown in Fig. 11. This load can model a light emitting diode (LED), and the boost converter is an LED driver. Assume that the output voltage variation *within* a cycle is small and ESR is also small, then the effective resistance of the CCL is close to $v_n/I_o$. The total load resistance (as a function of $v_n$) is $R_n = R(v_n) \approx R_0 \parallel (v_n/I_o) = R_0 v_n/(R_0 I_o + v_n)$. For $R_0 = \infty$, one has $R_n = v_n/I_o$ and the load is a pure CCL. For $I_o = 0$, one has $R_n = R_0$ and the load is a pure resistor.

From (20),

$$\Delta p_l = \frac{\rho T}{C}(\frac{1}{R} - \frac{1}{R_0}) = \frac{\rho T I_o}{vC} \tag{38}$$

For $I_o = 0$, one has $R = R_0$ and $\Delta p_l = 0$ because with a resistive load, the pole is not shifted by an additional term. If $I_o > 0$, then $\Delta p_l > 0$ and the pole is shifted to the right. If $I_o < 0$, then $\Delta p_l < 0$ and the pole is shifted to the left. The effect of CCL on the power stage or CMC is discussed next.

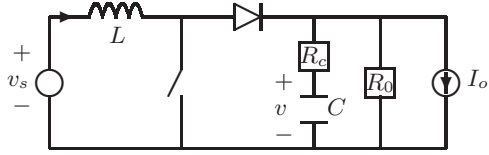

Figure 11: A boost converter power stage with CCL and resistive load $R_0$.

From (9) and (38), for the power stage or CMC,

$$p = p_0 + \Delta p_l = 1 - \frac{\rho T}{RC}(\frac{2M-1}{M-1}) + \frac{\rho T I_o}{vC} = 1 - \frac{\rho T}{RC}(\frac{M}{M-1}) - \frac{\rho T}{R_0 C} \quad (39)$$

Based on (39) and the fact that $1/R = 1/R_0 + I_o/v$, one can prove that for $I_o < v_s(1-2M)/R_0$, one has $p > 1$ and the converter is unstable with occurrence of saddle-node bifurcation. The occurrence of this condition may be rare because $R < 0$ is generally required. For $R > 0$ and $I_o < 0$ (or $R_0 > 0$), one has $p < 1$ and the saddle-node bifurcation does not occur.

### 7.2.2 Special Case: Pure CCL

Next, consider a special case that the load is a pure CCL ($R_0 = \infty$ and the total effective resistance $R = v/I_o$), for example. For $I_o > 0$, based on (39), one has $p < 1$ and the converter is generally stable (without occurrence of saddle-node bifurcation). For $I_o < 0$, based on (39), one has $p > 1$ and the converter is unstable. This result is reasonable because it agrees with the fact that, for $I_o < 0$, the capacitor voltage keeps increasing and there is no $T$-periodic orbit. Subharmonic oscillation (with $p < -1$) may occur if $M < 2RC/(2RC - \rho T)$, which is rare because another condition with $M > 1/(1-D)$ for the boost converter is required in DCM.

In the following two examples, the resistive load in Example 1 is replaced by a non-resistive load (but keeping the same effective resistance $R = 20$ Ω) to see the effects of different non-resistive loads.

In Example 4, the load is a pure CCL. With $I_o > 0$, the pole is shifted to the right. In Example 5, the load is a pure CCL in parallel with a resistor. With $I_o < 0$, the pole is shifted to the left. The other converter parameters remain the same as in Example 1.

**Example 4.** (*With $I_o > 0$, the pole is shifted to the right, confirmed with time simulation.*) Let the load be a pure CCL with $I_o = 0.9175$. From (3), $M = 3.67$. With $v = Mv_s = 18.35$, the effective resistance is $R = v/I_o = 20$ Ω. Based on the *exact* switching model, the exact pole is 0.9829, shifted from the open-loop pole 0.9707 by 0.0122.

**Accuracy of prediction.** The estimated pole from (39) is 0.9828, close to the exact pole 0.9829. The estimated pole shifting (38) due to CCL is $\Delta p_l = \rho T I_o/vC = 0.0125$, also agreed closely with the pole shifting 0.0122 based on the exact switching model.

**Verification by time-domain simulation.** A (cycle-by-cycle) simulation is made to confirm the pole location. Since $\hat{v}_{n+1} = p\hat{v}_n$, given an initial perturbation $\hat{v}_0$, at the end of $n$-th cycle, one has $\hat{v}_n = p^n \hat{v}_0$. The value of pole determines how fast $v_n$ converges to the fixed point $v$.

Let the circuit starts with an initial condition $(i_{L0}, v_0) = (0, 19)$ for a period of $20T$, shown in Fig. 12. The prediction of the discrete-time model (8) marked as * is close to the time plot. The circuit starts with an initial deviation $\hat{v}_0 = v_0 - v = 19 - 18.4175 = 0.5825$. The deviation (from the fixed point) $\hat{v}_n$ decays at a rate of $p = 0.9829$. At the end of the time period $20T$, the deviation becomes $\hat{v}_{20} = p^{20}\hat{v}_0 = 0.412$ and $v_{20} = 18.4175 + 0.412 = 18.83$, agreed with the time plot shown in Fig. 12. □

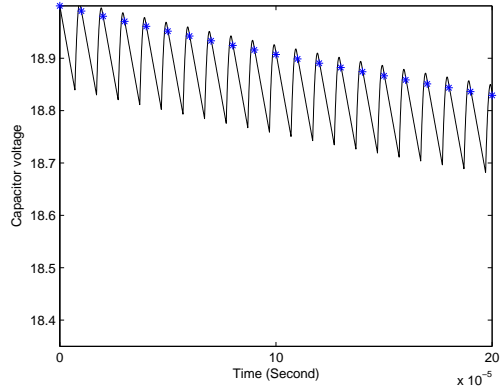

Figure 12: Time simulation of a boost converter with a pure CCL (with $I_o = 0.9175$). The discrete-time model prediction marked as * for $v_n$ is close to the time plot.

**Example 5.** (*With $I_o < 0$, the pole is shifted to the left, confirmed with time simulation.*) Let the load be a CCL with $I_o = -0.9175$ (negative) in parallel with a resistor $R_0 = 10\ \Omega$. With $v = Mv_s = 18.35$, the total effective resistance is $R_0 \parallel (v/I_o) = 20\ \Omega$. Based on the *exact* switching model, the exact pole is 0.9586, with a pole shifting (from the open-loop pole 0.9707) being -0.0121.

**Accuracy of prediction.** The estimated pole from (39) is 0.9578, close to the exact pole 0.9586. The estimated pole shifting (38) due to CCL is $\Delta p_l = \rho T I_o/vC = -0.0125$, also agreed closely with the pole shifting -0.0121 based on the exact switching model.

**Verification by time-domain simulation.** A time simulation is made to confirm the pole location. Let the circuit starts with an initial condition $(i_{L0}, v_0) = (0, 19)$ for a period of $20T$, shown in Fig. 13. The discrete-time model prediction marked as * is close to the time plot. The circuit starts with an initial deviation $\hat{v}_0$ and the deviation decays at a rate of $p = 0.9586$. Compared with Fig. 12, Fig. 13 shows a faster decay due to a smaller pole. □

## 8 Buck Converter

Generally, the pole shifting due to the non-resistive load is converter-independent, and the results derived above can be readily extended to the buck converter. The same pole shifting $\Delta p_l$ for the boost converter can be applied to the buck converter.

### 8.1 Scheme Four ($\mathbb{S}_4$): Buck Converter with a Pure CPL

#### 8.1.1 Open-Loop Power Stage

From (36) and (32), the power stage pole for the buck converter with a pure CPL (with the total effective resistance $R = v^2/P$) is

$$p = p_0 + \Delta p_l = 1 - \frac{\rho T}{RC}\left(\frac{M}{1-M}\right) = 1 - \frac{\rho T P}{v^2 C}\left(\frac{M}{1-M}\right) \tag{40}$$

which is stable [12] and agreed with [20, Eq. 17].

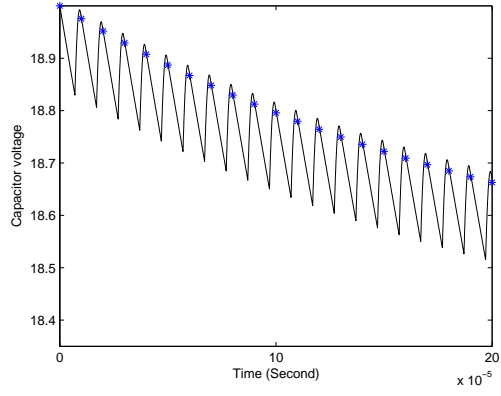

Figure 13: Time simulation of a boost converter with a CCL (with $I_o = -0.9175$) in parallel with a resistor $R_0 = 10$. The discrete-time model prediction marked as * for $v_n$ is close to the time plot.

### 8.1.2 Current Mode Control (CMC)

From (36) and (33), the pole for the CMC buck converter with a pure CPL is

$$p = p_0 + \Delta p_c + \Delta p_l = 1 - \frac{\rho T}{RC}(\frac{M - \frac{2M}{m_c}}{1 - M}) = 1 - \frac{\rho T P}{v^2 C}(\frac{M - \frac{2M}{m_c}}{1 - M}) \tag{41}$$

which is unstable (with $p > 1$) if no compensating ramp is added ($m_c = 1$). A ramp with $m_c > 2$ (or equivalently, $m_a > m_1$) is required to stabilize the converter.

## 8.2 Scheme Five ($\mathbb{S}_5$): Buck Converter with a Pure CCL

### 8.2.1 Open-Loop Power Stage

From (38) and (32), the power stage pole for the buck converter with a pure CCL (with the total effective resistance $R = v/I_o$) is

$$p = p_0 + \Delta p_l = 1 - \frac{\rho T}{RC}(\frac{1}{1 - M}) = 1 - \frac{\rho T I_o}{vC}(\frac{1}{1 - M}) \tag{42}$$

For $I_o > 0$, one has $p < 1$ and the converter is generally stable. For $I_o < 0$, one has $p > 1$ and the converter is unstable. Subharmonic oscillation (with $p < -1$) may occur if $M > 1 - \rho T/2RC$ (close to 1), which is rare because another condition [10] with $M < D$ for the buck converter is required in DCM.

### 8.2.2 Current Mode Control (CMC)

From (38) and (33), the pole for the CMC buck converter with a pure CCL is

$$p = p_0 + \Delta p_c + \Delta p_l = 1 - \frac{\rho T}{RC}(\frac{1 - \frac{2M}{m_c}}{1 - M}) = 1 - \frac{\rho T I_o}{vC}(\frac{1 - \frac{2M}{m_c}}{1 - M}) \tag{43}$$

which is unstable with $p > 1$, for example, if $I_o > 0$, $M > 1/2$ and no compensating ramp is added ($m_c = 1$). Saddle-node bifurcation occurs when $M = 1/2$. The existence of saddle-node bifurcation can be proved in another way based on steady-state analysis discussed next.

Table 3: Poles for boost and buck converters with pure CPL or CCL.

|  | Boost converter | Buck converter |
|---|---|---|
| **Pue CPL** ($R_0 = \infty$), $\Delta p_l = \frac{2\rho T}{RC}$ | | |
| Power stage | $1 - \frac{\rho T}{RC}(\frac{1}{M-1})$ | $1 - \frac{\rho T}{RC}(\frac{M}{1-M})$ |
| CMC | (same as above) | $1 - \frac{\rho T}{RC}(\frac{M - \frac{2M}{m_c}}{1-M})$ |
| | | unstable if $m_c < 2$ |
| **Pue CCL** ($R_0 = \infty$), $\Delta p_l = \frac{\rho T}{RC}$ | | |
| Power stage | $1 - \frac{\rho T}{RC}(\frac{M}{M-1})$ | $1 - \frac{\rho T}{RC}(\frac{1}{1-M})$ |
| CMC | (same as above) | $1 - \frac{\rho T}{RC}(\frac{1 - \frac{2M}{m_c}}{1-M})$ |
| | | unstable if $m_c < 2M$ |
| (Poles may be also unstable if $R < 0$.) | | |

From (31), in steady state (to determine the fixed points), let $D_n = D$, $v_{sn} = v_s$ and $v_{n+1} = v_n = v = Mv_s$. Then (31) leads to a steady-state equation,

$$M^2 - M + \frac{v_c^2 L}{2TI_o v_s} = 0 \tag{44}$$

which is a quadratic equation of $M$ and has two solutions,

$$M = \frac{1}{2} \pm \sqrt{\frac{1}{4} - \frac{v_c^2 L}{2TI_o v_s}} \tag{45}$$

The stable solution (fixed point) has $M < 1/2$ and the unstable solution has $M > 1/2$. From (45), the two solutions coalesce and thus the saddle-node bifurcation occurs when $M = 1/2$ and

$$v_c = \sqrt{\frac{TI_o v_s}{2L}} \tag{46}$$

From (43), a ramp with $m_c > 2M$ (or equivalently, $m_a > (2M - 1)m_1$) is required to stabilize the converter.

A summary of poles for boost and buck converters with pure CPL or CCL ($R_0 = \infty$) is given in Table 3.

**Example 6.** (*Saddle-node bifurcation with coexistence of a stable orbit and an unstable orbit, confirmed with time simulation.*) Consider a CMC buck converter with $v_s = 5$ V, $f_s = 200$ kHz, $L = 5$ $\mu$H, $C = 40$ $\mu$F, and $v_c = 0.9$ A. The load is a pure CCL with $I_o = 0.4$ A.

**Results based on the exact switching model.** Based on the *exact* switching model, there coexist two $T$-periodic orbits shown in Fig. 14 in the time domain and Fig. 15 in state space. The two orbits have the same peak inductor current set by $v_c = 0.9$. They also have the same average inductor current 0.4, which equals to $I_o$. The first orbit has $M = 0.28$, $v = Mv_s = 1.4$ and its pole is 0.9785, which is stable. The second orbit has $M = 0.72$, $v = Mv_s = 3.6$ and its pole is 1.022, which is unstable.

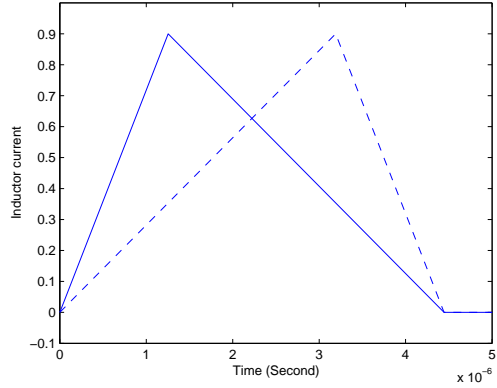

Figure 14: *Coexistence* of a stable orbit (solid line) and an unstable orbit (dashed line) in *time domain*, $v_c = 0.9$.

**Results based on the discrete-time model.** From (43), the poles for these two orbits are 0.9785 and 1.022, respectively, agreed exactly with the exact switching model. If $v_c = 1$, the two orbits (now with $M = 1/2$) coalesce and the saddle-node bifurcation occurs, predicted exactly by (46).

**Simulation verification.** Two time simulations are made to verify the stability and instability of the poles. First, let the circuit starts with an initial condition $(i_{L0}, v_0) = (0, 1.5)$ for a period of $200T$, shown in Fig. 16. The circuit starts with an initial deviation $\hat{v}_0 = 0.15 - 0.14 = 0.1$ and the deviation decays at a rate of $p = 0.9785$. At the end of the time period $200T$, the deviation becomes $\hat{v}_{200} = p^{200} \hat{v}_0 = 0.0013$, and the state trajectory of the capacitor voltage converges toward the expected stable fixed point $v = M v_s = 1.4$.

Next, let the circuit starts with an initial condition $(i_{L0}, v_0) = (0, 3.7)$ for a period of $50T$, shown in Fig. 17. The circuit starts with an initial deviation $\hat{v}_0 = 3.7 - 3.6 = 0.1$ and the deviation grows at a rate of $p = 1.0215$. At the end of the time period $50T$, the deviation grows to $\hat{v}_{50} = p^{50} \hat{v}_0 = 0.029$, and the state trajectory of the capacitor voltage moves away from the unstable fixed point $v = M v_s = 3.6$.

These two time simulations confirm the stability and instability of the two orbits shown in Fig. 15. Similar to the discussion in [13], if the circuit starts with an initial condition with $i_L = 0$ and $v < 3.6$, the state trajectory will converge toward the stable fixed point (with $v = 1.4$). If the circuit starts with an initial condition with $i_L = 0$ and $v > 3.6$, the state trajectory will move away from the unstable fixed point (with $v = 3.6$). □

# Part III
# Variable Frequency Control

Similar analysis can be applied to variable frequency control, such as valley voltage constant-on-time control (denoted here as V-COTC) [25] or boundary (borderline) conduction mode (BCM) [16, 17, 21]. The BCM is also called critical conduction mode. Variable switching frequency is generally applied for ripple-based control [26] where only simple *static* feedback (instead of high-order *dynamic* feedback) is required.

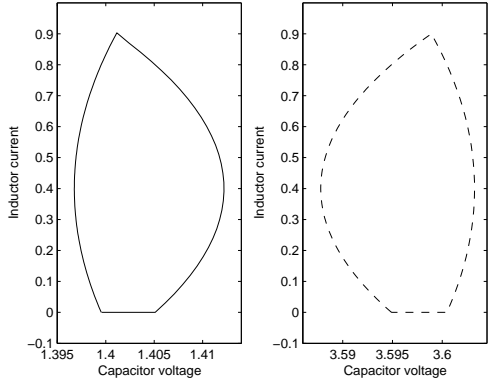

Figure 15: *Coexistence* of a stable orbit (solid line) and an unstable orbit (dashed line) in *state space*, $v_c = 0.9$.

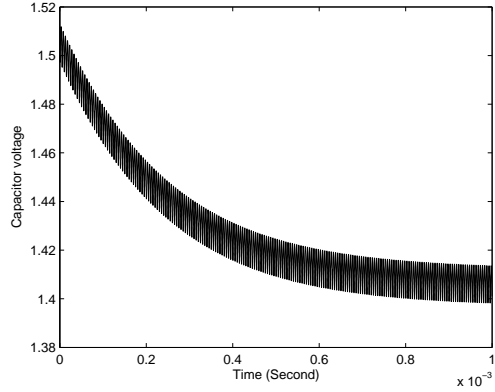

Figure 16: The circuit starts with an initial condition $(i_{L0}, v_0) = (0, 1.5)$ for a period of $200T$. The state trajectory of the capacitor voltage converges toward the expected stable fixed point $v = Mv_s = 1.4$.

In the variable frequency control, the $n$-th cycle period $T_n$ is not fixed. An illustrative signal plot of $i_L$ is shown in Fig. 18. There are two feedback variables $d_n$ and $T_n$. Without loss of generality, only a resistive load is considered and hence $R_n = R$. With a non-resistive load, the pole is just shifted by $\Delta p_l$ as discussed in Sec. 7.

## 9 Scheme Six ($\mathbb{S}_6$): Valley Voltage Constant-On-Time Control (V-COTC)

A V-COTC boost converter is shown in Fig. 20. For either a buck converter or a boost converter, the operation of V-COTC is as follows. In a variable cycle period $T_n$ which has three stages, the switch is turned on in the first stage for a fixed time duration $d = DT$ (hence $d_n = d$ and $D_n = D$), the switch is turned off and the diode is turned on in the second stage (for a time duration $D_2T$), and the switch is turned on again at $t = \sum_{i=1}^{n} T_i$ (or at $t = T_n$ *within* the cycle) when $v_o$ drops below a control signal $v_c$ plus a compensating ramp with a slope $m_a$. This control scheme puts a

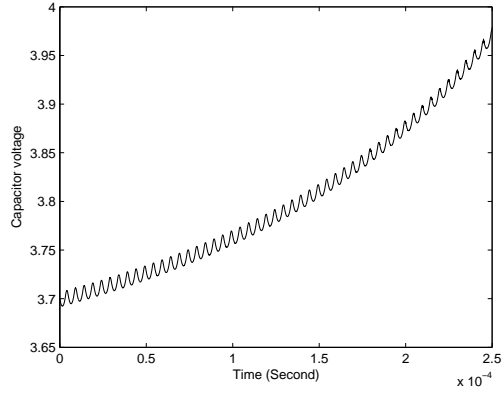

Figure 17: The circuit starts with an initial condition $(i_{L0}, v_0) = (0, 3.7)$ for a period of $50T$. The state trajectory of the capacitor voltage moves away from the unstable fixed point $v = Mv_s = 3.6$.

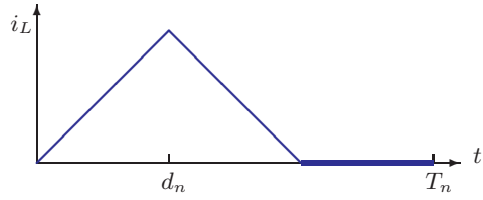

Figure 18: An illustrative signal plot of $i_L$ in each cycle for DCM under variable frequency control.

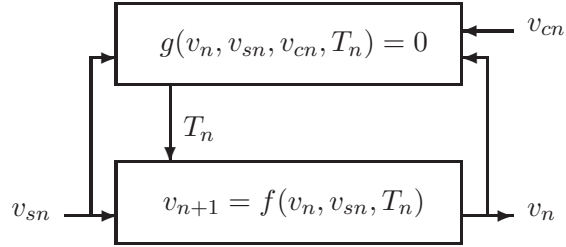

Figure 19: Closed-loop large-signal dynamics for V-COTC.

constraint equation [22] at the end of the $n$-th cycle (at the switching instant $t = \sum_{i=1}^{n} T_i$):

$$g(v_n, v_{sn}, v_{cn}, T_n) = v_{n+1} - v_{cn} - m_a T_n = 0 \tag{47}$$

From (1), the power stage dynamics is

$$v_{n+1} = f(v_n, v_{sn}, T_n) = (1 - \frac{\rho T_n}{RC})v_n + \frac{\rho d^2 v_{sn}^2}{2LC(v_n - v_{sn})} \tag{48}$$

The closed-loop dynamics is shown in Fig. 19.

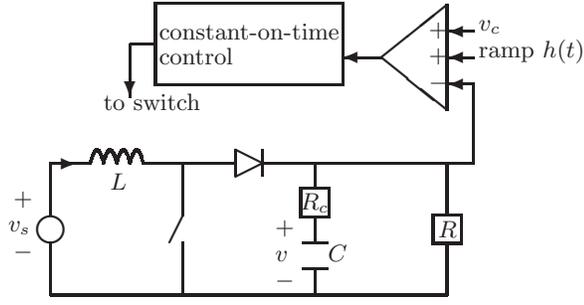

Figure 20: A boost converter under V-COTC.

The closed-loop pole (also valid for the buck converter) is

$$p = \frac{\partial f}{\partial v_n} - \frac{\partial f}{\partial T_n}(\frac{\partial g}{\partial T_n})^{-1}\frac{\partial g}{\partial v_n} \tag{49}$$

$$= p_0 - (\frac{\frac{\rho M v_s}{RC}}{\frac{\rho M v_s}{RC} + m_a})p_0 \tag{50}$$

$$:= p_0 + \Delta p_c \tag{51}$$

Without the ramp ($m_a = 0$), one has $\Delta p_c = -p_0$ and $p = 0$, resulting a deadbeat effect [10]. Adding the ramp removes the deadbeat effect.

A converter generally has two dynamic variables associated with the inductor current and the capacitor voltage, and hence has two poles. In DCM, the pole associated with the inductor current is zero. With the constant-on-time control, the second pole associated with the capacitor voltage is also zero, agreed with [9].

## 10   Scheme Seven ($\mathbb{S}_7$): Boundary Conduction Mode (BCM)

The BCM can be considered as a special case of DCM and CCM. An illustrative signal plot of $i_L$ is shown in Fig. 21. The operation of BCM is as follows. *Within* the $n$-th cycle, the switch is turned on for a period $d_n$, and the switch is turned off at $t = T_n$ when $i_L = 0$. The cycle period $T_n$ is the sum of the on-time ($d_n$) and the off-time ($T_n - d_n$). By simple algebra based on the inductor current slopes for the on-time and the off-time, the cycle period $T_n$ for BCM can be proved to be proportional to the on-time $d_n$. For the boost converter,

$$T_n = \frac{d_n v_n}{v_n - v_{sn}} \tag{52}$$

For the buck converter,

$$T_n = \frac{d_n v_{sn}}{v_n} \tag{53}$$

Thus, $T_n$ is a function of $d_n$ and it can be *internalized* into the closed-loop dynamics (shown later).

Two BCM schemes are considered. In the first scheme, the first stage has variable on-time $d_n$. The switch is turned off at $t = d_n$ *within* a cycle when $i_L = v_{cn}$. This control scheme puts a constraint equation (similar to (25)) at the switching instant $t = d_n$ within the cycle. For the boost converter,

$$d_n = \frac{v_{cn}L}{v_{sn}} \tag{54}$$

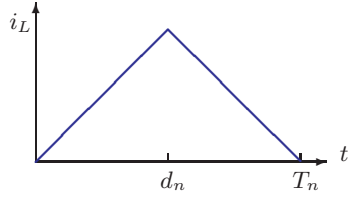

Figure 21: An illustrative signal plot of $i_L$ in each cycle for BCM.

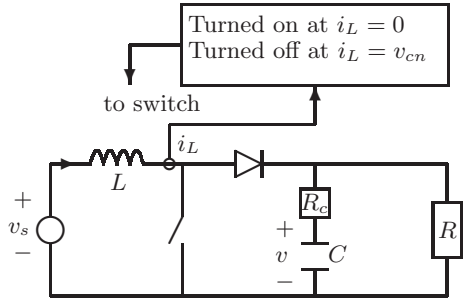

Figure 22: A boost converter under BCM.

For the buck converter,

$$d_n = \frac{v_{cn}L}{v_{sn} - v_n} \tag{55}$$

In the second scheme, the first stage has constant on-time $d_n = d = DT$, instead of variable on-time. For simplicity, the first scheme is denoted here simply as BCM, and the second scheme is denoted as BCM-COT.

Since the BCM can be considered as a special case of DCM and CCM, the steady-state equations for DCM and CCM can be applied in BCM. The following steady-state equations are used throughout the paper to simplify the small-signal dynamics. From (3) and (4), for the boost converter in DCM, one has $D = \sqrt{KM(M-1)}$. For the boost converter in BCM (operating at the CCM-DCM boundary), $K = D(1-D)^2 = (M-1)/M^3$. For the buck converter in DCM, from [10, p. 124], $D = \sqrt{KM^2/(1-M)}$. For the buck converter in BCM, $K = 1 - D = 1 - M$.

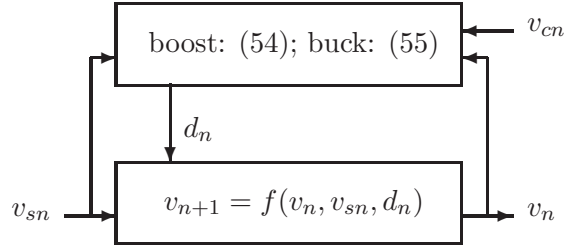

Figure 23: Closed-loop large-signal dynamics for BCM.

## 10.1 Boost Converter

### 10.1.1 BCM with Variable On-Time

A BCM boost converter is shown in Fig. 22. From (52), similar to (48), the large-signal dynamics is

$$v_{n+1} = f(v_n, v_{sn}, d_n) = (1 - \frac{\rho d_n v_n}{RC(v_n - v_{sn})})v_n + \frac{\rho d_n^2 v_{sn}^2}{2LC(v_n - v_{sn})} \quad (56)$$

where the constraint equation for $d_n$ is shown in (54). The dynamics is shown in Fig. 23, where the dynamic variable $T_n$ is internalized.

By taking partial derivatives of (56), the small-signal dynamics is

$$\hat{v}_{n+1} = (1 - \frac{2\rho T}{RC})\hat{v}_n + \frac{\rho T M}{RC}\hat{v}_{sn} + \frac{\rho T}{2CM}\hat{v}_{cn} \quad (57)$$

The pole is stable and its location agrees with [16] by a simple continuous-to-discrete mapping. From (57), the DC gain of control-to-output transfer function is

$$T_{oc}(1) = \frac{R}{4M} \quad (58)$$

agreed with [21]. Similarly from (57), the DC gain of audio-susceptibility is

$$T_{os}(1) = \frac{M}{2} \quad (59)$$

also agreed with [16].

### 10.1.2 BCM with Constant On-Time (BCM-COT)

A BCM-COT boost converter is shown in Fig. 25. By setting $d_n = d = DT$ in (56), the large-signal dynamics is

$$v_{n+1} = f(v_n, v_{sn}) = (1 - \frac{\rho d v_n}{RC(v_n - v_{sn})})v_n + \frac{\rho d^2 v_{sn}^2}{2LC(v_n - v_{sn})} \quad (60)$$

The dynamics is shown in Fig. 24, where the dynamic variable $T_n$ is internalized.

By taking partial derivatives of the large-signal dynamics, the small-signal dynamics is

$$\hat{v}_{n+1} = (1 - \frac{2\rho T}{RC})\hat{v}_n + \frac{\rho T}{RC}(\frac{2M^2}{M-1})\hat{v}_{sn} \quad (61)$$

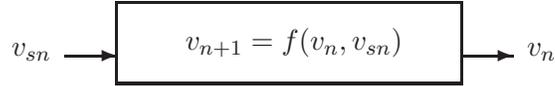

Figure 24: Large-signal dynamics for BCM-COT.

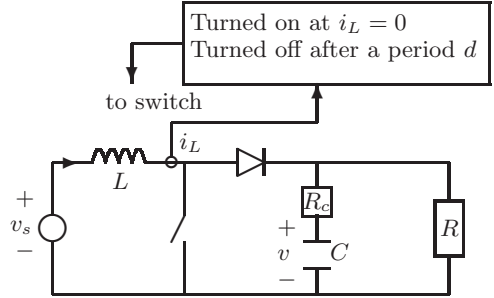

Figure 25: A boost converter under BCM-COT.

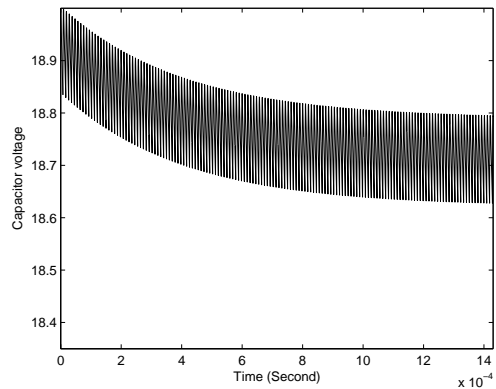

Figure 26: The circuit starts with an initial condition $(i_{L0}, v_0) = (0, 19)$ for 150 cycles. The *sampled* state trajectory of the capacitor voltage $v_n$ converges toward $v = 18.787$.

**Example 7.** (*The discrete-time model accurately predicts the pole location, confirmed with time simulation.*) Consider a BCM-COT boost converter with parameters $v_s = 5$ V, $R = 20$ $\Omega$, $L = 5$ $\mu$H, $C = 40$ $\mu$F, $R_c = 0$, and $d = 7$ $\mu$s.

A time simulation is made to confirm the pole location. Let the circuit starts with an initial condition $(i_{L0}, v_0) = (0, 19)$ for 150 cycles (with a variable period in each cycle), shown in Fig. 26. The circuit starts with an initial deviation $\hat{v}_0 = 0.213$ and the deviation decays at a rate of $p = 0.975$. At the end of the 150 cycles, the deviation becomes $\hat{v}_{150} = p^{150}\hat{v}_0 = 0.0048$, and the *sampled* state trajectory of the capacitor voltage $v_n$ converges toward the fixed point $v = 18.787$. □

## 10.2 Buck Converter

### 10.2.1 BCM with Variable On-Time

From (53), similar to (31), the large-signal dynamics for the BCM buck converter is

$$v_{n+1} = (1 - \frac{\rho d_n v_{sn}}{RC v_n})v_n - \frac{\rho d_n^2 v_{sn}}{2LC}(1 - \frac{v_{sn}}{v_n}) \tag{62}$$

where the constraint equation for $d_n$ is shown in (55).

By taking partial derivatives of (62), the small-signal dynamics is

$$\hat{v}_{n+1} = (1 - \frac{\rho T}{RC})\hat{v}_n + \frac{\rho T}{2C}\hat{v}_{cn} \tag{63}$$

The pole is stable and its location agrees with [16]. From (63), the DC gain of control-to-output transfer function is

$$T_{oc}(1) = \frac{R}{2} \tag{64}$$

agreed with [16]. Similarly from (63) which does not have a term associated with $\hat{v}_{sn}$, the audio-susceptibility is zero, also agreed with [16].

### 10.2.2 BCM with Constant On-Time (BCM-COT)

By setting $d_n = d = DT$ in (62), the large-signal dynamics for the BCM-COT buck converter is

$$v_{n+1} = (1 - \frac{\rho d v_{sn}}{RC v_n})v_n - \frac{\rho d^2 v_{sn}}{2LC}(1 - \frac{v_{sn}}{v_n}) \tag{65}$$

By taking partial derivatives of the large-signal dynamics, the small-signal dynamics is

$$\hat{v}_{n+1} = (1 - \frac{\rho T^2}{2LC})\hat{v}_n + \frac{\rho T}{RC}(\frac{M}{1 - M})\hat{v}_{sn} \tag{66}$$

A summary for boost and buck converters with variable frequency control is given in Table 4.

# 11 Conclusion

A simple one-dimensional discrete-time model is used to analyze the boost converter in discontinuous conduction mode. Boost and buck converters in seven different schemes are analyzed systematically. The linearized dynamics is obtained simply by taking partial derivatives with respect to dynamic variables. The discrete-time model provides a simpler alternative (different from the circuit-averaging approach) to design or analyze the converter. The key results are summarized in Tables 2-4.

In the discrete-time model, there is only a single pole and no zero. The single closed-loop pole is a linear combination of three terms: the open-loop pole, a term due to the control scheme, and a term due to the non-resistive load. Even with a single pole, the phase response of the discrete-time model can go beyond -90 degrees as in the two-pole average models.

In the boost converter with a resistive load under current mode control, adding the compensating ramp has no effect on the pole location. Increasing the ramp slope decreases the DC gain of control-to-output transfer function and increases the audio-susceptibility. The ramp, beneficial in CCM to stabilize the current loop, may be unnecessary in DCM since the current loop is not oscillatory.

Table 4: Summary for boost and buck converters with variable frequency control, some agreed with past research results [16, 21].

|  | Boost converter |  | Buck converter |  |
|---|---|---|---|---|
| **V-COTC** |  |  |  |  |
| Pole, $p$ | $(\frac{m_a}{m_a - \frac{\rho M v_s}{RC}})p_0$ |  | $(\frac{m_a}{m_a - \frac{\rho M v_s}{RC}})p_0$ |  |
| Pole, $m_a = 0$ (no ramp) | 0, dead-beat |  | 0, dead-beat |  |
| **BCM** |  |  |  |  |
| Pole, $p$ | $1 - \frac{2\rho T}{RC}$ | [16] | $1 - \frac{\rho T}{RC}$ | [16] |
| DC gain, $T_{oc}(1)$ | $\frac{R}{4M}$ | [21] | $\frac{R}{2}$ | [16] |
| Audio-susceptibility, $T_{os}(1)$ | $\frac{M}{2}$ | [16] | 0 | [16] |
| **BCM-COT** |  |  |  |  |
| Pole, $p$ | $1 - \frac{2\rho T}{RC}$ |  | $1 - \frac{\rho T^2}{2LC}$ |  |
| Audio-susceptibility, $T_{os}(1)$ | $\frac{M^2}{M-1}$ |  | $\frac{KM}{1-M}$ |  |

Similar analysis is applied to the buck converter with a non-resistive load or variable switching frequency. The derived dynamics agrees closely with the exact switching model and the past research results. Similar analysis can be applied to other types of converters in DCM. The derived discrete-time models can be also applied to other applications, such as power factor correction and *digital* control of DC-DC converters.

# References

[1] S. Ćuk and R. D. Middlebrook, "A general unified approach to modelling switching DC-to-DC converters in discontinuous conduction mode," in *Proc. IEEE PESC*, 1977, pp. 36–57.

[2] A. S. Kislovski, R. Redl, and N. O. Sokal, *Dynamic Analysis of Switching-Mode DC/DC Converters*. New York: Van Nostrand Reinhold, 1991.

[3] D. Maksimovic and S. Ćuk, "A unified analysis of PWM converters in discontinuous modes," *IEEE Trans. Power Electron.*, vol. 6, no. 3, pp. 476–490, 1991.

[4] V. Vorperian, "Simplified analysis of PWM converters using model of PWM switch. II. Discontinuous conduction mode," *IEEE Trans. Aerosp. Electron. Syst.*, vol. 26, no. 3, pp. 497–505, 1990.

[5] J. Sun, D. M. Mitchell, M. F. Greuel, P. T. Krein, and R. M. Bass, "Averaged modeling of PWM converters operating in discontinuous conduction mode," *IEEE Trans. Power Electron.*, vol. 16, no. 4, pp. 482–492, 2001.

[6] A. Reatti and M. K. Kazimierczuk, "Small-signal model of PWM converters for discontinuous conduction mode and its application for boost converter," *IEEE Trans. Circuits Syst. I*, vol. 50, no. 1, pp. 65–73, 2003.

[7] A. Davoudi, J. Jatskevich, and T. De Rybel, "Numerical state-space average-value modeling of PWM DC-DC converters operating in DCM and CCM," *IEEE Trans. Power Electron.*, vol. 21, no. 4, pp. 1003–1012, 2006.

[8] C.-C. Fang, "Bifurcation boundary conditions for current programmed PWM DC-DC converters at light loading," *Int. J. of Electron.*, 2011, accepted and published online, DOI:10.1080/00207217.2012.669715.

[9] R. B. Ridley, "A new continuous-time model for current-mode control with constant on-time, constant off-time, and discontinuous conduction mode," in *Proc. IEEE PESC*, 1990, pp. 382–389.

[10] R. W. Erickson and D. Maksimovic, *Fundamentals of Power Electronics*, 2nd ed. Berlin, Germany: Springer, 2001.

[11] M. K. Kazimierczuk, *Pulse-width Modulated DC-DC Power Converters*. Wiley, 2008.

[12] A. M. Rahimi and A. Emadi, "Discontinuous-conduction mode DC/DC converters feeding constant-power loads," *IEEE Trans. Ind. Electron.*, vol. 57, no. 4, pp. 1318–1329, 2010.

[13] C.-C. Fang, "Unified discrete-time modeling of buck converter in discontinuous mode," *IEEE Trans. Power Electron.*, vol. 26, no. 8, pp. 2335–2342, 2011.

[14] C. K. Tse and K. M. Adams, "Qualitative analysis and control of a DC-to-DC boost converter operating in discontinuous mode," *IEEE Trans. Power Electron.*, vol. 5, no. 3, pp. 323–330, 1990.

[15] C. K. Tse, "Flip bifurcation and chaos in three-state boost switching regulators," *IEEE Trans. Circuits Syst. I*, vol. 41, no. 1, pp. 16–23, 1994.

[16] J. Chen, R. Erickson, and D. Maksimovic, "Averaged switch modeling of boundary conduction mode DC-to-DC converters," in *Proceedings of IEEE IECON*, 2001, pp. 844–849.

[17] T. Suntio, "Average and small-signal modeling of self-oscillating flyback converter with applied switching delay," *IEEE Trans. Power Electron.*, vol. 21, no. 2, pp. 479–486, 2006.

[18] C.-C. Fang, "Saddle-node bifurcation in the buck converter with constant current load," *Nonlinear Dynamics*, vol. 69, no. 4, pp. 1739–1750, 2012.

[19] ——, "Sampled data poles, zeros, and modeling for current mode control," *Int. J. of Circuit Theory Appl.*, 2011, accepted and published online, DOI: 10.1002/cta.790.

[20] V. Grigore, J. Hatonen, J. Kyyra, and T. Suntio, "Dynamics of a buck converter with a constant power load," in *Proc. IEEE PESC*, 1998, pp. 72–78.

[21] B. Choi, S.-S. Hong, and H. Park, "Modeling and small-signal analysis of controlled on-time boost power-factor-correction circuit," *IEEE Trans. Ind. Electron.*, vol. 48, no. 1, pp. 136–142, 2001.

[22] C.-C. Fang and E. H. Abed, "Sampled-data modeling and analysis of power stages of PWM DC-DC converters," *Int. J. of Electron.*, vol. 88, no. 3, pp. 347–369, March 2001.

[23] ——, "Saddle-node bifurcation and Neimark bifurcation in PWM DC-DC converters," in *Nonlinear Phenomena in Power Electronics: Bifurcations, Chaos, Control, and Applications*, S. Banerjee and G. C. Verghese, Eds. New York: Wiley, 2001, pp. 229–240.

[24] C.-C. Fang, "Sampled-data poles and zeros of buck and boost converters," in *Proc. IEEE Int. Symp. Circuits Sys.*, vol. 3, Orlando, FL, USA, 1999, pp. 731–734.

[25] J. Li and F. C. Lee, "Modeling of V$^2$ current-mode control," *IEEE Trans. Circuits Syst. I*, vol. 57, no. 9, pp. 2552–2563, 2010.

[26] R. Redl and J. Sun, "Ripple-based control of switching regulators - an overview," *IEEE Trans. Power Electron.*, vol. 24, no. 12, pp. 2669–2680, Dec. 2009.